\documentclass[onecolumn,a4paper,longbibliography,accepted=2022-03-22]{quantumarticle}
\pdfoutput=1
\usepackage[T1]{fontenc}
\usepackage[latin9]{inputenc}
\usepackage{amsmath}
\usepackage{amssymb}
\usepackage{graphicx}
\usepackage{babel}
\usepackage{hyperref}

\usepackage[numbers]{natbib}
\usepackage{authblk}

\begin{document}

\global\long\def\ket#1{\left|#1\right\rangle }%

\global\long\def\bra#1{\left\langle #1\right|}%

\global\long\def\braket#1#2{\left\langle #1\left|#2\right.\right\rangle }%

\global\long\def\ketbra#1#2{\left|#1\right\rangle \left\langle #2\right|}%

\global\long\def\braOket#1#2#3{\left\langle #1\left|#2\right|#3\right\rangle }%

\global\long\def\mc#1{\mathcal{#1}}%

\global\long\def\nrm#1{\left\Vert #1\right\Vert }%

\title{Bounds on the recurrence probability in periodically -driven quantum
systems}

\author[huji]{Tanmoy Pandit}
\author[umd]{Alaina M. Green}
\author[umd]{C. Huerta Alderete}
\author[umd]{Norbert M. Linke}
\author[huji]{Raam Uzdin}
\affil[huji]{Fritz Haber Research Center for Molecular Dynamics, Hebrew
University of Jerusalem, Jerusalem 9190401, Israel}
\affil[umd]{Joint Quantum Institute and Department of Physics, University of Maryland, College Park, MD 20742 USA}

\email{tanmoy.pandit@huji.ac.il, raam@mail.huji.ac.il}

\begin{abstract}
Periodically-driven systems are ubiquitous in science and technology.
In quantum dynamics, even a small number of periodically-driven spins
leads to complicated dynamics. Hence, it is of interest to understand
what constraints such dynamics must satisfy. We derive
a set of constraints for each number of cycles. For pure initial states,
the observable being constrained is the recurrence probability. We
use our constraints for detecting undesired coupling to unaccounted environments
and drifts in the driving parameters. To illustrate
the relevance of these results for modern quantum systems we demonstrate
our findings experimentally on a trapped-ion quantum computer, and
on various IBM quantum computers. Specifically, we provide two experimental
examples where these constraints surpass fundamental bounds associated
with known one-cycle constraints. This scheme can potentially be used to detect
the effect of the environment in quantum circuits that cannot be classically
simulated. Finally, we show that, in practice, testing an $n$-cycle
constraint requires executing only $O(\sqrt{n})$ cycles, which makes
the evaluation of constraints associated with hundreds of cycles realistic.
\end{abstract}
\maketitle
The dynamics of isolated quantum systems become too hard to simulate on a classical computer or solve analytically when more than a few dozen spins are involved. Yet, the unitary evolution that characterizes isolated quantum systems is not arbitrary. It conserves purity, the von Neumann entropy, and any other function that depends on the eigenvalues of the density matrix.
However, these conserved quantities are impractical since state tomography or other non-scalable techniques are typically needed to evaluate them. Here we study isolated quantum systems under periodic driving for two reasons. First, to gain a fundamental understanding of how the unitarity of isolated quantum systems restricts the evolution of observables, and second, to show the relevance of these restrictions to noise detection in state-of-the-art quantum computers.
We start from the question: is the evolution of observables in a periodically-driven system more restricted compared to non-periodic driving? While is seems intuitively true, to the best of our knowledge no such constraints have been shown. We derive such constraints and show that
these inequalities are more restrictive than other known bounds, which
do not exploit periodicity. In Sec. \ref{sec:Periodicity-inequalities} we introduce
and derive the periodicity inequalities, and in section \ref{sec: Experimental-demonstrations}
we experimentally demonstrate their added value by exploring pure
and mixed-state scenarios that are challenging for existing bounds
as described below. 


Constraints on observables that are valid for arbitrary driving have
been derived within the frameworks of stochastic and quantum thermodynamics.
These constraints include, the second law in quantum microscopic setups
fluctuation theorems \cite{harris2007fluctuation,seifert2012stochastic,jarzynski2011equalities},
thermodynamic uncertainty relation \cite{polettini2016tightening,timpanaro2019thermodynamic,koyuk2018generalization,barato2015thermodynamic}, passivity based constraints \cite{uzdin2018global,uzdin2021passivity,strasberg2021first},
and more. We do not include constraints that involve non-observable,
information-like, quantities as in resource theory and other frameworks
\cite{strasberg2021first,bera2019thermodynamics}. As it turns out,
thermodynamic constraints loose their predictive power in the zero-temperature limit \cite{timpanaro2020landauer,jarzynski2006rare,uzdin2018second}.
In appendix I, we show this explicitly for the second law and the
Jarzynski fluctuation theorem. However, in quantum computers and simulators
the input state is typically pure, which makes many thermodynamic-based
constraints unsuitable for evolution-noise detection. Another challenge
for thermodynamic-like constraints involves coherence in the energy
basis. Fluctuation theorems, for example, are based on the two-point
measurement scheme. The first measurement is carried on the initial
state, which leads to wavefunction collapse. Crucially this measurement
changes the evolution from a unitary evolution to a mixture of unitaries.

We show the added value of our inequalities by: i) demonstrating that
the periodicity bounds can detect violation of periodicity even if
the system is isolated; ii) showing that the periodicity constraints
remains useful for detecting unaccounted coupling to an environment,
in the zero-temperature temperature limit; iii) breaking a recently 
derived bound on the maximal temperature of detectable environments.\\

\section*{Preliminaries: A Simple Inequality For Unitary Evolution}

Let $\rho_{f}$ be the density matrix that is obtained from $\rho_{0}$ by unitary transformation. Due to the non-negativity of the L2 norm it holds that, 
\begin{eqnarray*}
tr[(\rho_{0}-\rho_{f})^{2}] & \ge & 0,
\end{eqnarray*}
Next, we expand this expression and use the cyclic property of the trace to obtain the following form

\begin{eqnarray}
tr[\rho_{0}^{2}]+tr[\rho_{f}^{2}]-2tr[\rho_{0}\rho_{f}]\ge0,
\end{eqnarray}

Since the evolution  $\rho_{0}\to\rho_{f}$  is unitary,is unitary,the purity is
conserved, $tr[\rho_{0}^{2}]=tr[\rho_{f}^{2}]$, and we obtain

\begin{eqnarray}
tr[\rho_{0}(\rho_{0}-\rho_{f})]\ge0.\label{eq: F x GP}
\end{eqnarray}

Next, we use the fact that for unitary evolution purity conservation can be generalized to any trace value   $tr[(\rho_{0}-aI)^{2}]=tr[(\rho_{f}-aI)^{2}]$ where $I$ is
the identity operator and $a$ is any real number. The proof follows
trivially by expending the parentheses. Thus, if $r_0$ is some Hermitian
operator (potentially traceless) and  evolves unitarily to $r_f$, i.e.
$r_{f}=Ur_{0}U^{\dagger}$ where $U$ is a unitary matrix, then by repeating
the steps that lead to (\ref{eq: F x GP}) together with $tr[r_{0}^{2}]=tr[r_{f}^{2}]$
we obtain
\begin{eqnarray}
tr[r_{0}(r_{0}-r_{f})]\ge0.\label{eq: r0rf}
\end{eqnarray}

\section{Periodicity inequalities\label{sec:Periodicity-inequalities}}

\subsection{The two-cycle inequality}

Next, we consider a periodically-driven system, i.e. the density matrix
after each cycle satisfies 
\begin{eqnarray}
\rho_{k+1}=U\rho_{k}U^{\dagger},\label{eq: periodicity}
\end{eqnarray}
where $U$ is a unitary evolution operator. The fact that $U$ does
not depend on $k$ reflects the assumption of periodic driving. From
linearity, it follows that the object 
\begin{eqnarray}
r_{0}=\sum_{j=0}^{N}\alpha_{j}\rho_{j},
\end{eqnarray}
where $\alpha_{j}\in\mathbb{R}$ and $\rho_{j}$ is the density matrix
after j cycles, also satisfies $r_{k+1}=Ur_{k}U^{\dagger}$. 
In the following, we use the term `stencil' for $r$. According
to (\ref{eq: r0rf}) it holds that 
\begin{eqnarray}
tr[r_{0}(r_{0}-r_{M})]\ge0,\label{eq: gen periodic GP}
\end{eqnarray}
where $r_{M}=U^Mr_{0}U^{\dagger M}$.Consider the simple case where $r_{0}=\rho_{1}-\rho_{0}$, and $r_{M}=r_{1}=\rho_{2}-\rho_{1}$,
the inequality (\ref{eq: gen periodic GP}) leads to
\begin{eqnarray}
0 & \le & tr[(\rho_{1}-\rho_{0})(-\rho_{0}+2\rho_{1}-\rho_{2})]\nonumber \\
 & = & tr[\rho_{0}^{2}]+2tr[\rho_{1}^{2}]-3tr[\rho_{0}\rho_{1}]\nonumber \\
 &  & -tr[\rho_{1}\rho_{2}]+tr[\rho_{0}\rho_{2}].
\end{eqnarray}
$R_{k}=tr[\rho_{0}\rho_{k}]$ is the key quantity in
our work and we shall refer to it as the \emph{recurrence probability}.
For a pure state, $\rho_{0}$ is the projector on to the initial state,
$\rho_{0}=\ketbra{\phi_{0}}{\phi_{0}}$. Therefore, $tr[\rho_{0}\rho_{k}]$
is the probability to measure the system in the initial state $\ket{\phi_{0}}$
after $k$ cycles. $R_{k}$ is an inner product, and can therefore
be thought of as the amount of overlap with the initial state even
for mixed $\rho_{0}$. That is, $R_{k}$ still expresses to what extent
the density matrix returns to its initial value after $k$ cycles. 

Using purity conservation in unitary evolution, $tr[\rho_{0}^{2}]=tr[\rho_{1}^{2}]$,
it holds that
\begin{eqnarray}
3R_{0}-3R_{1}-tr[\rho_{1}\rho_{2}]+R_{2}\ge0.\label{eq: temp three point}
\end{eqnarray}
However, $-tr[\rho_{1}\rho_{2}]$ is presently in an inconvenient
form as both $\rho_{1}$ and $\rho_{2}$ are unknown. To overcome
this we use the fact that the driving is periodic,
\begin{eqnarray}
tr[\rho_{j}\rho_{j+k}] & = & tr[\rho_{j}U^{j}\rho_{k}U^{\dagger j}]\nonumber \\
 & = & tr[U^{\dagger j}\rho_{j}U^{j}\rho_{k}]\nonumber \\
 & = & tr[\rho_{0}\rho_{k}]=R_{k}.\label{eq: invariance unital}
\end{eqnarray}
Applying eq. (\ref{eq: invariance unital}) in eq. (\ref{eq: temp three point})
we get
\begin{eqnarray}
R_{2}-4R_{1}+3R_{0}\ge0,\label{eq: 3 point}
\end{eqnarray}
which is the simplest inequality in our approach since it contains
only two cycles, or three time points including the initial state.
Periodicity was assumed for the driving and the density matrix does
not evolve periodically in the general case. Note that the coefficients
in (\ref{eq: 3 point}) sum up to zero. When the evolution is the identity operator,so that $R_k=1$, this property leads 
to a saturation of the inequality \ref{eq: 3 point} of the form $0\ge0$.
\begin{figure}
\includegraphics[width=11cm]{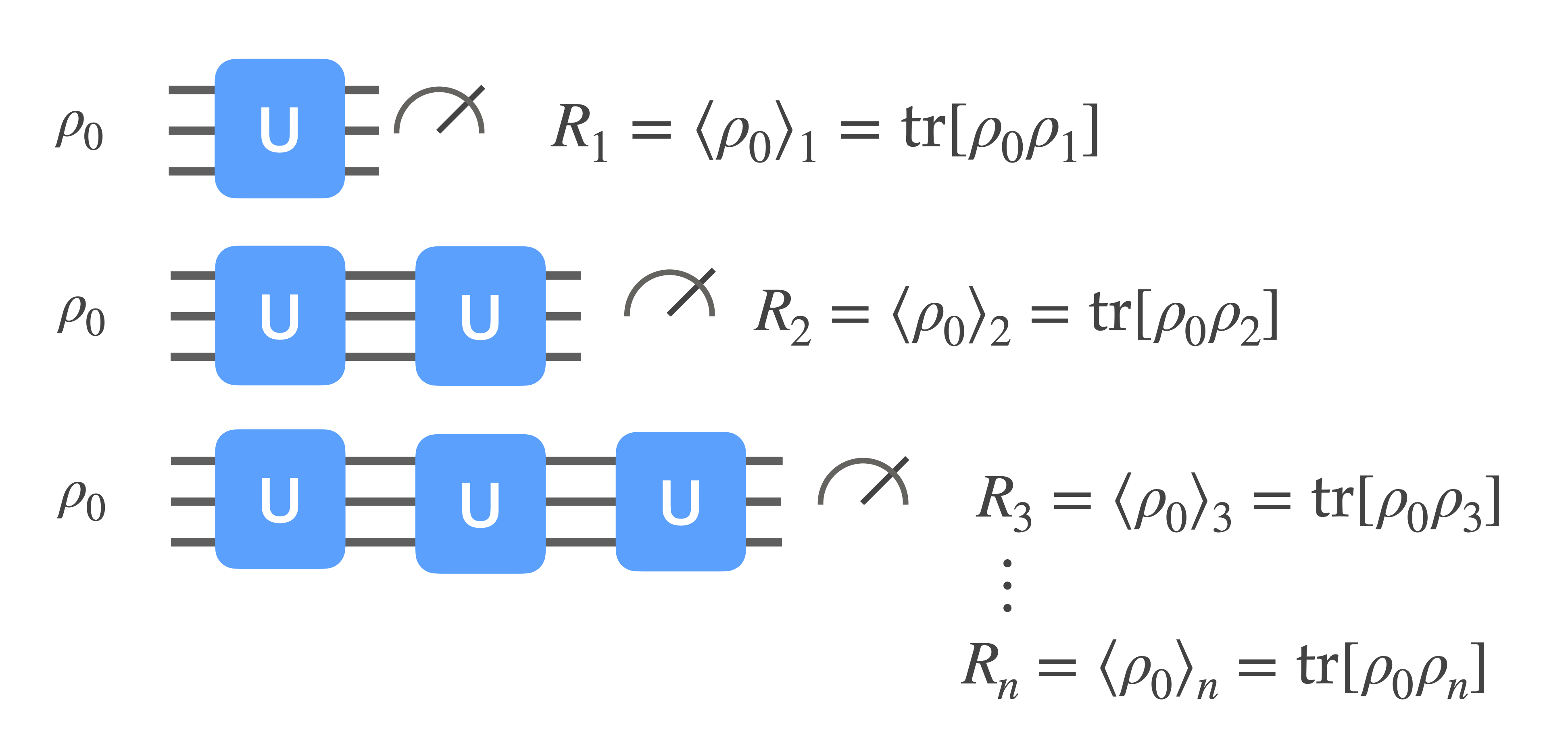}

\caption{\label{fig: illust}Illustration of the protocol for obtaining the
$k$ cycle recurrence probability $R_{k}$ which appears in the periodicity
inequalities {[}eqn. (\ref{eq: S2})-(\ref{eq: S4}) etc.{]}. The
recurrence probability $R_{k}$ expresses to what extent the system
has returned to its initial state after $k$ cycles. The different
$R_{k}$ are statistically independent, and the measurement of $R_{k}$
do not affect $R_{j>k}$ since they are evaluated in different experiments.
Hence, despite the quantum evolution and the multiple measurements
in this protocol, there is no wavefunction ``collapse'' that modifies
the evolution.}

\end{figure}

Operationally, the inequality in (\ref{eq: 3 point}) is evaluated
as illustrated in Fig. \ref{fig: illust}. The $R_{0}$ measurement
can be skipped if the state preparation and measurement errors (SPAM)
are negligible. After the initial state is evolved for one cycle,
$R_{1}$ is measured. Next, in a third set of measurements, $R_{2}$
is measured, and so on. 

Note that the one-cycle inequality $R_{0}\ge R_{1}$ becomes trivial
for pure states since $R_{0}=1$ and $1\ge R_{1}$ always holds as
 the recurrence probability cannot exceed one. In contrast, (\ref{eq: 3 point})
provides a non-trivial prediction for pure $\rho_{0}$, which is a
typical starting point for quantum devices. In Appendix II, we further
discuss the two-cycle inequality and its added value for mixed states.

\subsection{\label{subsec:Multi-time-point-inequalities}A recipe for constructing
multi-cycle inequalities}

The following recipe can be used to obtain more general periodicity
bounds:
\begin{itemize}
\item Choose a stencil of the form $r_{0}=\sum_{j=0}^{N}\alpha_{j}\rho_{j}$
($\alpha_{i}$ are real numbers).
\item Choose $M$ to set the shift value in $r_M=U^M r_0$. 
\item Expand the parenthesis in $tr[r_{0}(r_{0}-r_{M})]\ge0$. 
\item Replace $tr[\rho_{m}\rho_{m+k}]\to R_{k}=tr[\rho_{0}\rho_{k}]$.
\end{itemize}
The last stage includes the $m=0$ case: $tr[\rho_{k}^{2}]\to R_{0}=tr[\rho_{0}^{2}]$. The number of cycles is determinied by largest index of $R_k$ that appears in $r_M$. Thus the number of cycles in the resulting inequality is $n=N+M$.

\subsection{The discrete derivative inequalities $S_{n}$\label{subsec:The-discrete-derivative}}

The two-cycle inequality (\ref{eq: 3 point}) was obtained by using the stencil $r_0=\rho_1-\rho_0$ which has the form of a first-order discrete time derivative. Next, we exploit higher-order discrete derivative to generate additional stencils. We denote by $r_{0}^{(n)}$ the shifted discrete derivative $r_{0}^{(1)}=\rho_{1}-\rho_{0},r_{0}^{(2)}=\rho_{2}-2\rho_{1}+\rho_{0},...$
The second derivative $r_{0}^{(2)}$ is centered at $k=1$, The third
derivative is centered at $k=2$ and so on. This shift ensures that the derivative does not include negative
indices. For $r_{M}$ we choose $r_{1}$. Applying the recipe in Sec.
(\ref{subsec:Multi-time-point-inequalities}) to the stencils $r_0=(\frac{1}{2})^{n}r_{0}^{(n-1)}$,
and denoting the resulting inequalities by $S_{n}$ where $n$ is
the number of cycles, the first few inequalities are
\begin{eqnarray}
S_{2}=\frac{1}{8}(R_{2}-4R_{1}+3R_{0})\ge0,\label{eq: S2}\\
S_{3}=\frac{1}{32}(-R_{3}+6R_{2}-15R_{1}+10R_{0})\ge0,\\
S_{4}=\frac{1}{128}(R_{4}-8R_{3}+28R_{2}-56R_{1}+35R_{0})\ge0.\label{eq: S4}
\end{eqnarray}
The $S_{n}$ inequalities can be written as $S_{n}=\sum_{k=0}^{n}w_{k}^{(n)}R_{k}\ge0$
where the coefficients $w_{k}^{(n)}$ alternate sign as a function
of $k$ and satisfy $\sum_{k=0}^{n}w_{k}^{(n)}=0$ and $\sum_{k=0}^{n}\text{\ensuremath{\left|w_{k}^{(n)}\right|}}=1$.
As shown later, when $n$ is sufficiently large the coefficients take
the form: $w_{k}^{(n)}\to\frac{(-1)^{k}}{\sqrt{\pi n}}e^{-\frac{k^{2}}{n}}$.

\subsection{\label{subsec:The-scaling-law}Low-cost evaluation of the $S_{n}$
periodicity inequalities}

We introduce the notation $A_{\pm}\rho_{n}=\rho_{n\pm1}$, and show in Appendix III that the $S_n$ inequalities can be written as  
\begin{eqnarray}
S_{n}=tr[\rho_{0}(\frac{1}{2}-\frac{1}{4}A_{-}-\frac{1}{4}A_{+})^{n}\rho_{0}].\label{eq: SnA}
\end{eqnarray}
The expression $(\frac{1}{2}-\frac{1}{4}A_{-}-\frac{1}{4}A_{+})^{n}$
describes a classical $n$-step random walk with probability $1/4$
to move one step forward, $1/4$ to move one step backward, and $1/2$
to stay in place. For large $n$, the probability distribution for
moving moving $k$ steps forward is given by the normal distribution
with single-step variance $\sigma_{0}^{2}$:
\begin{eqnarray}
\frac{1}{\sqrt{2\pi\sigma_{0}^{2}n}}e^{-\frac{k^{2}}{2\sigma_{0}^{2}n}}=\frac{1}{\sqrt{\pi n}}e^{-\frac{k^{2}}{n}},
\end{eqnarray}
where we have used: $\sigma_{0}^{2}=\frac{1}{2}\cdot0+2\cdot\frac{1}{4}\cdot1=\frac{1}{2}$.
Applying this result to eq. (\ref{eq: SnA}) we find
\begin{eqnarray}
S_{n} & = & \sum_{k=-n}^{n}\frac{(-1)^{k}}{\sqrt{\pi n}}e^{-\frac{k^{2}}{n}}R_{\left|k\right|}\nonumber \\
 & = & \frac{1}{\sqrt{\pi n}}R_{0}+2\sum_{k=1}^{n}\frac{(-1)^{k}}{\sqrt{\pi n}}e^{-\frac{k^{2}}{n-1}}R_{k}.\nonumber \\
\label{eq: Sn gaussian}
\end{eqnarray}
Expression (\ref{eq: Sn gaussian}) suggests that the coefficients
$w_{k}^{(n)}$ decay very fast as a function of $k$ and therefore
the series $S_{n}$ can potentially be truncated at some value $L$
without significantly changing the value of $S_{n}$. Operationally,
it means that fewer than $n$ cycles are needed for evaluating
$S_{n}$. This is a useful feature for detecting small evolution noise, that may require evaluating $S_{n}$ with large $n$.

Next, we set a bound on the error incurred by truncating the sum at $L<n$. 

\begin{eqnarray}
\sum_{k=L}^{n}\frac{(-1)^{k}}{\sqrt{\pi n}}e^{-\frac{k^{2}}{n}}R_{k} & \le & \sum_{k=L}^{n}\frac{1}{\sqrt{\pi n}}e^{-\frac{k^{2}}{n}}\nonumber \\
 & \le & \sum_{k=L}^{\infty}\frac{1}{\sqrt{\pi n}}e^{-\frac{k^{2}}{n}}\nonumber \\
 & \simeq & +\frac{1}{2}\text{erfc}(\frac{L}{\sqrt{n}})\frac{1}{2\sqrt{\pi n}}e^{-\frac{L^{2}}{n}}.
\end{eqnarray}
For convenience we set $L=\xi\sqrt{n}$ and get
\begin{eqnarray}
2\sum_{k=L}^{n}\frac{1}{\sqrt{\pi n}}e^{-\frac{k^{2}}{n}}\simeq\text{erfc}(\xi)+\frac{1}{\sqrt{\pi n}}e^{-\xi^{2}}.\label{eq: erfc}
\end{eqnarray}
From (\ref{eq: Sn gaussian}) and (\ref{eq: erfc}) we finally obtain
that $S_{n}^{(L)}$ the $L$-cycle truncated version of the $S_{n}$
satisfies
\begin{eqnarray}
S_{n}^{(L=\xi\sqrt{n})} & = & S_{n}-\sum_{k=L}^{n}w_{k}R_{k}\nonumber \\
 & \ge & -2\sum_{k=L}^{n}(-1)^{k}\frac{1}{\sqrt{\pi n}}e^{-\frac{k^{2}}{n}}R_{k}\nonumber \\
 & \ge & -\text{erfc}(\xi)-\frac{1}{\sqrt{\pi n}}e^{-\xi^{2}}.\label{eq: S_N^M}
\end{eqnarray}
This inequality should be used as follows: $n$ is fixed by the $S_{n}$
that needs to be evaluated, the required accuracy $\epsilon$ is
used to determine $\xi$ via $\epsilon=\text{erfc}(\xi)+\frac{1}{\sqrt{\pi n}}e^{-\xi^{2}}$
and finally, the truncation value is $L=\xi\sqrt{n}$. For example,
to get an error of $5\times10^{-3}$ in the calculation of $S_{1000}$
($1000$ cycles), only $L=64$ cycles need to be measured ($\xi=2$).
For an accuracy of $10^{-5}$, $99$ cycles are needed ($\xi=\pi$),
and $126$ measured cycles already yield an accuracy of $2\times10^{-8}$.
We emphasize that in practice the accuracy is limited by statistical
noise. For example in the experiment in section \ref{subsec: Realistic-heat-leak}
the truncation error is $1/65$ of the $3\sigma$-width when taking
$24$ points ($\xi=2.1$). For $30$ points ($\xi=2.63$), the truncation
error is already $1/1000$ of the $3\sigma$ width.
\subsection{Optimized three-cycle inequality}

In the following, we explore the continuum of three-cycle inequalities,
i.e. we are not using the discrete derivative stencil as in the previous
sections. The goal is produce bounds that can perform better at detecting
evolution noise. Consider the general three-cycle stencil,
\begin{eqnarray}
r_{0}=\rho_{0}+x\rho_{1}+y\rho_{2},\label{eq: ab stencil}
\end{eqnarray}
where $x$ and $y$ can take any real value. Using the recipe in Sec.
\ref{subsec:Multi-time-point-inequalities} we get the following inequality:
\begin{eqnarray}
L(x,y) & = & -yR_{3}+(-x+2y-xy)R_{2}\nonumber \\
 & + & (-1+2x-y-(x-y)^{2})R_{1}\nonumber \\
 & + & (1-x+x^{2}+y^{2}-xy)R_{0}\ge 0. \label{eq: Lab}
\end{eqnarray}
We notice that the left hand side, which we denote by $L(x,y)$, is
a second-order polynomial in $x$ and $y$, where the coefficients
of $x,y,xy,x^{2},y^{2},1$ depend on $\{R_{n}\}_{n=0}^{3}$. 

If the evolution is unitary and periodic, $L(x,y)\ge0$ for any value
of $x$ and $y$. Yet, in the presence of evolution noise, $L(x,y)$
might be positive for some choices of $x$ and $y$ and negative for
other choices. It is also possible that $L(x,y)$ is not negative
for any choice of $x$ and $y$. In that case, the evolution noise
is not detectable with four-point periodicity inequalities. To test
if the noise is detectable we check if $\underset{x,y}{\text{min}}[L(x,y)]<0$ by finding the optimal $x$ and $y$ that minimize L.

In a given physical scenario that consists of an initial condition
and a driving protocol, $\{R_{n}\}_{n=0}^{3}$ are just constant positive
numbers and $L(x,y)$ is a second-order polynomial with known and
fixed coefficients. Next we check which type of paraboloid $L(x,y)$
is. Evaluating the second-derivative test for a paraboloid $\partial_{x}^{2}L\partial_{y}^{2}L-(\partial_{x}\partial_{y}L)^{2}\ge0$,
we find $\partial_{x}^{2}L\partial_{y}^{2}L-(\partial_{x}\partial_{y}L)^{2}=(R_{0}-R_{1})(3R_{0}-4R_{1}+R_{2})=8(R_{0}-R_{1})S_{2}$.
If $S_{2}<0$ it means that $L(x,y)$ is hyperbolic paraboloid and
for some value of $x,y$ it must be negative. $L(x,y)<0$ implies detection, but this is not surprising since the noise is already
detectable with two cycles since $S_2<0$. $S_2$ corresponds to $x=-1$, $y=0$. The more interesting case is $S_{2}>0$
where $L(x,y)$ is a paraboloid. Now detection $L(x,y)<0$ is not
guaranteed. Furthermore, the motivation for using an optimized bound
for evolution noise detection is that $S_{2}$ and $S_{3}$, which
use $\{R_{n}\}_{n=0}^{3}$, failed to detect it. To determine if $L(x,y)$
is a convex or concave paraboloid we check the limit $x\to\infty,y=0$.
Since $L(x,y)$ is positive in this limit we conclude $L(x,y)$ is
convex. Consequently, the most negative value of $L(x,y)$ occurs
when $\partial_{x}L(x,y)=0$ and $\partial_{y}L(x,y)=0$, we denote
the solution to these two equations by $x_{min}$ and $y_{min}$.
Finding $x_{min},y_{min}$ we get that $L(x_{min},y_{min})\ge0$ is
equal to
\begin{eqnarray}
\frac{(2R_{0}-R_{1}-2R_{2}+R_{3})}{(R_{0}-R_{2})(3R_{0}-4R_{1}+R_{2})}\left(R_{0}^{2}-R_{0}(R_{1}+R_{3})-(R_{1}-R_{2})^{2}+R_{1}R_{3}\right) & \ge & 0.\label{eq: raw opt4pnt}
\end{eqnarray}

This bound is non-linear in the expectation values $R_{k}$ since these 
expectation values are being used to choose the optimal $x$ and $y$
for detection. If the inequalities that involve $R_{0},R_{1}$ and
$R_{2}$ hold, i.e. $R_{0}-R_{2}\ge0$ and $3R_{0}-4R_{1}+R_{2}\ge0$
, then the non-linear bound (\ref{eq: raw opt4pnt}) simplifies
to
\begin{eqnarray}
R_{0}^{2}-R_{0}(R_{1}+R_{3})-(R_{1}-R_{2})^{2}+R_{1}R_{3}\ge0.\label{eq: opt 4pnt}
\end{eqnarray}
The non-linearity in this inequality is different from the density
matrix non-linearity that is usually considered in quantum information and
quantum thermodynamics (e.g. purity, von Neumann entropy $S_{vn}=-tr[\rho\ln\rho]$,
etc.) as here it appears outside the trace. An experimental measurement
of this inequality is given in Sec. \ref{subsec:Noise-Detection hot}. 

While optimized bounds become more burdensome to derive as the number
of cycles increases, the existence of these bounds can enable the
detection of violations with a relatively small number of measurements. 
\section{Numerical and Experimental demonstrations\label{sec: Experimental-demonstrations}}

We present numerical simulations and experimental results from IBM
superconducting quantum processors and a trapped ion quantum computer
(TIQC). Each example illustrates a different feature of the new bounds.

\subsection{Detecting parameter drift during the evolution}

There are only two assumptions made in deriving the periodicity inequalities, unitarity and  periodic driving. A violation
of the periodicity inequalities can arise from breaking either one
or both of these assumptions. In this section we confirm that a non-periodic shift in the control parameters would manifest as a violation of these bounds by numerically investigating such a possibility.

\begin{figure}
\includegraphics[width=12cm]{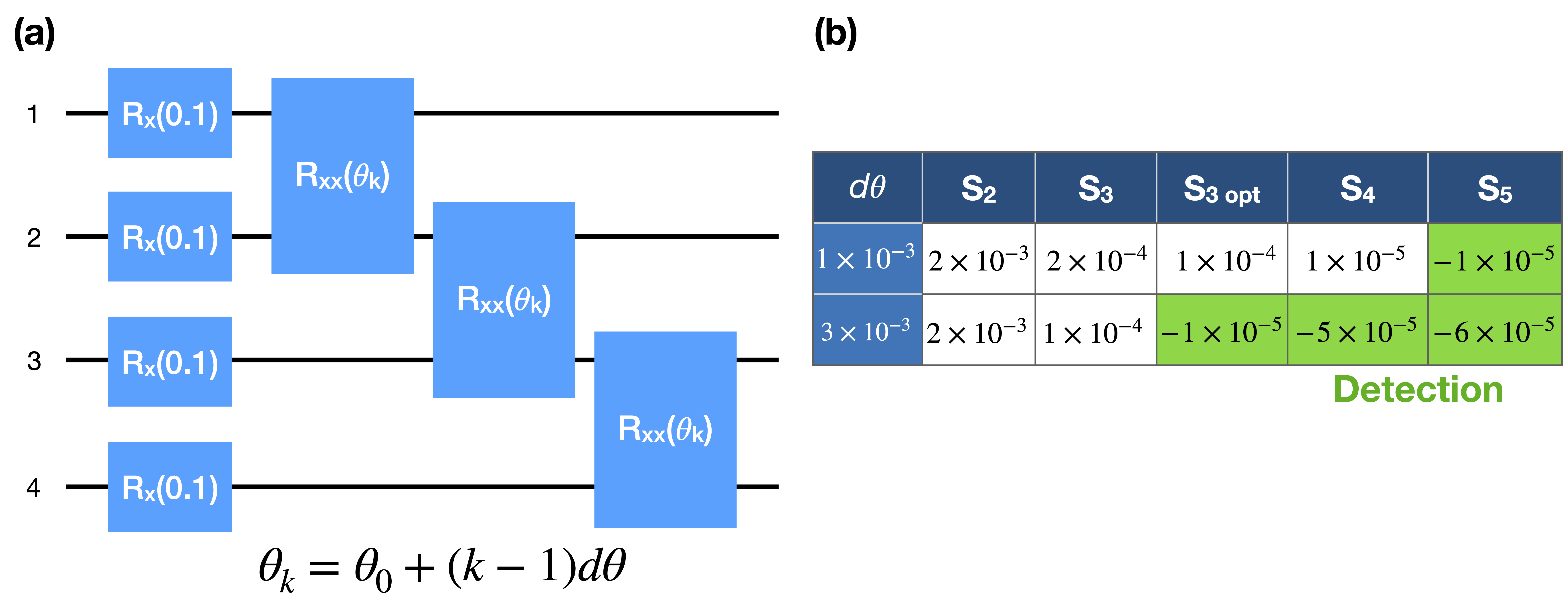}

\caption{\label{fig: nonPeriodic}(a) In this circuit the $\sigma_{x}\sigma_{x}$
interaction ($R_{xx}$ ) changes form one cycle to the next ($k$
dependence). (b) For a small drift of $d\theta=10^{-3}$ the non periodicity
is detectable after five cycles, i.e. $S_{5}<0$. When $d\theta=3\times10^{-3}$,
dour cycle are enough for detecting the non-periodicity. Interestingly,
it is possible to detect the non-periodicity by running only three
cycles by applying the optimized 3-cycle periodicity inequality (\ref{eq: opt 4pnt}).
To the best of our knowledge, the periodicity inequalities are the
first constraints on unitary dynamics that are customized for periodic
driving and as such the detection of non-periodicity is presently
unique to these inequalities. }

\end{figure}

In Fig. \ref{fig: nonPeriodic}a we numerically study a four-qubit
system where the driving cycle is composed of four single-qubit rotations
around the x-axis by angle of 0.1rad $R_{x}(0.1)$ where $R_x(\theta)=e^{-i \frac{1}{2}\theta \sigma}$ is a single qubit rotation around the x-axis of the Bloch sphere. These rotations
are followed by nearest neighbor $xx$ interactions between qubit
$j$ and qubit $j+1$ that leads to the following evolution: $U_{k}^{[j,j+1]}=\exp[i\frac{\theta_{k}}{2}(\sigma_{x}^{(i)}\sigma_{x}^{(i+1)}]$.
In order to simulate a drift of the control parameters, we chose $\theta_{k}=\theta_{0}+(k-1)d\theta$.
$d\theta$ represents the drift of the parameter $\theta$ from one
cycle to the next. The circuit is repeated for up to five cycles.
Choosing $\theta_{0}=0.3$ rad and $d\theta=10^{-3}rad$, the first
violation is observed after five cycles where $S_{5}=-1.56\times10^{-5}$.
If the drift is stronger, for example such that $d\theta=3\times10^{-3}rad$,
the first violation is observed after three cycles when employing
the three-cycle optimized inequality (\ref{eq: opt 4pnt}), which  yields
the value $S_{3\:opt}=-1.5\times10^{-5}$. The unoptimized three cycle
bound yields $S_{3}=+1.4\times10^{-4}.$ Alternatively, an additional
cycle can be used to detect the drift without resorting to an optimized
bound $S_{4}=-5.1\times10^{-5}$. Other values appear in Fig. \ref{fig: nonPeriodic}b.

The violation of the periodicity inequality shown in this example
highlights the power of these inequalities over thermodynamic bounds, 
which could only detect errors arising from non-unitarity. In this
case, we detect the drift error in the device by exploiting tighter
bounds arising from the periodicity of the intended evolution. 
\subsection{Evolution noise detection with a pure initial state}

\begin{figure}
\includegraphics[width=11cm]{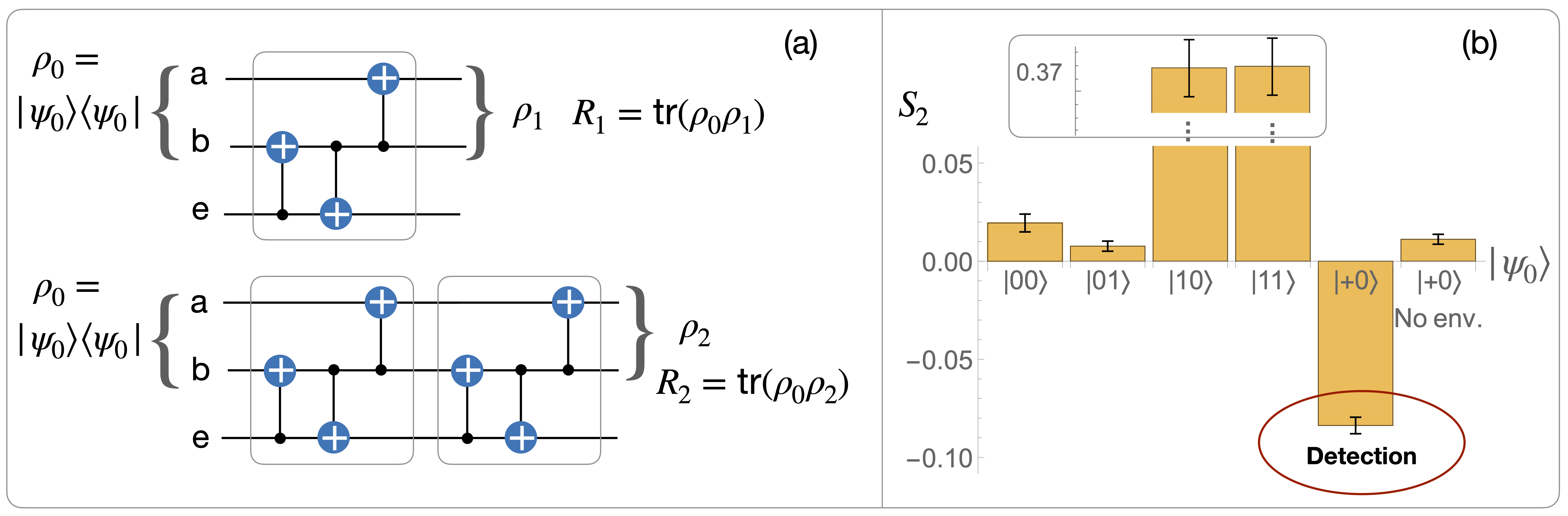}

\caption{\label{fig: ctrl hl pure}(a) a circuit for demonstrating heat leak
detection with a pure initial state. The initial condition is $\protect\ket{\psi_{0}}\protect\bra{\psi_{0}}$
in the upper two qubits which constitute the system and the bottom
environment qubit is initially in $\protect\ket 0\protect\bra 0$.
The recurrence probability of the two upper spins is measured after
one cycle in one experiment, and after two cycles in a different experiment.
(b) The experimental values of $S_{2}$ for various input states carried
out on the IBM Bogota processor. The negative value for $\protect\ket{\psi_{0}}=\protect\ket{+_{a}0_{b}}$
indicates the detection of the environment. When the environment is
decoupled for the same input state (``No env.'' in the figure) $S_{2}$
yields a positive value as expected. This experiment illustrates detection
of evolution noise using pure state which is impossible in thermodynamically-inspired frameworks. }
\end{figure}

In some setups, such as digital quantum computers, pure-state initial
conditions are more natural to use compared to mixed states, which
require some dedicated preparation protocol (e.g. see \cite{henao2019experimental}
and \cite{zhu2020generation}). It is natural, then, to
ask if pure states can be used to detect evolution noise. Presently,
constraints on observables in unitary dynamics are derived within
the frameworks of stochastic and quantum thermodynamics. Thus, they
are the relevant reference for the periodicity inequalities we present.
Unfortunately, thermodynamic constraints become impractical to use
when the initial state is sufficiently pure. As discussed in Appendix
I, some thermodynamic inequalities become so loose that no evolution
noise can violate them. In other thermodynamic constraints, the resources
needed for evaluating and measuring them diverge as the
initial state becomes pure. Hence, thermodynamic constraints
are presently not suitable for detecting evolution noise in quantum computers
and simulators. The periodicity inequalities studied here are free
from this limitation. We present a proof-of-principle experiment for
evolution noise detection using pure states. Our setup, shown in Figure
\ref{fig: ctrl hl pure}(a), is composed of a two-qubit system coupled
to a one-qubit environment ('e'). The initial state is created by
 single-qubit gates, which are not shown in the figure. The single
cnot gate between the upper two qubits constitutes the periodic unitary
driving. While this environment is artificial, it has the advantage
that we can easily decouple it from the system and verify that there is 
no noise detection in this case.

The experiment is composed of three sub-experiments for measuring
$R_{0},R_{1}$ and $R_{2}$. We test several different input states:
$\ket{00},\ket{01},\ket{10},\ket{11}$, and $\ket{1+}$. Our goal
is to observe a violation of $S_{2}$ when the environment is connected,
and no violation when it is not. Figure \ref{fig: ctrl hl pure}(b)
shows that states in the computational basis cannot detect the evolution
noise but the superposition state in the computation basis $\ket{1+}$
can. The error bars correspond to $\pm3\sigma$ statistical uncertainty.
The last bar in Figure \ref{fig: ctrl hl pure}(b) shows that, as
expected, when the environment qubit is decoupled $S_{2}$ is positive
and there is no detection. That is, we confirm that our inequality
does not always yield negative values regardless of the coupling to
the environment qubit.

Interestingly, the circuit is completely classical for binary input,
i.e. diagonal states in the computational basis. In this case, it acts 
  as a simple permutation in the computation basis $0_{a}1_{b}\leftrightarrow1_{a}1_{b}$.
The positive values for all four elements of the computation basis {[}Fig. \ref{fig: ctrl hl pure}(b){]} imply that  no classical binary input (including stochastic inputs with
classical correlations) can be used to detect the noise induced by
the environment qubit 'e', in this
circuit. Yet, a quantum superposition of two classical
binary states can detect the coupling to the environment. While this
example does not involve a strong quantum feature, such as entanglement,
it naturally raises the question if entangled states can lead to detection
of noise that cannot be detected using classically-correlated states.
The key point of this experiment is the demonstration of noise detection
using a unitary constraint when starting in a pure state. A task that,
to the best of our knowledge, is not possible using the known thermodynamic
constraints. 

To conclude this section we point out that while a violation indicates a deviation from a perfect periodic unitary driving, isolating the cause such as parameter drift, decoherence, spontaneous emission, or other evolution noise mechanisms, requires further tests.

\subsection{\label{subsec: Realistic-heat-leak}Detection of realistic noise}

\begin{figure}
\begin{centering}
\includegraphics[width=15cm]{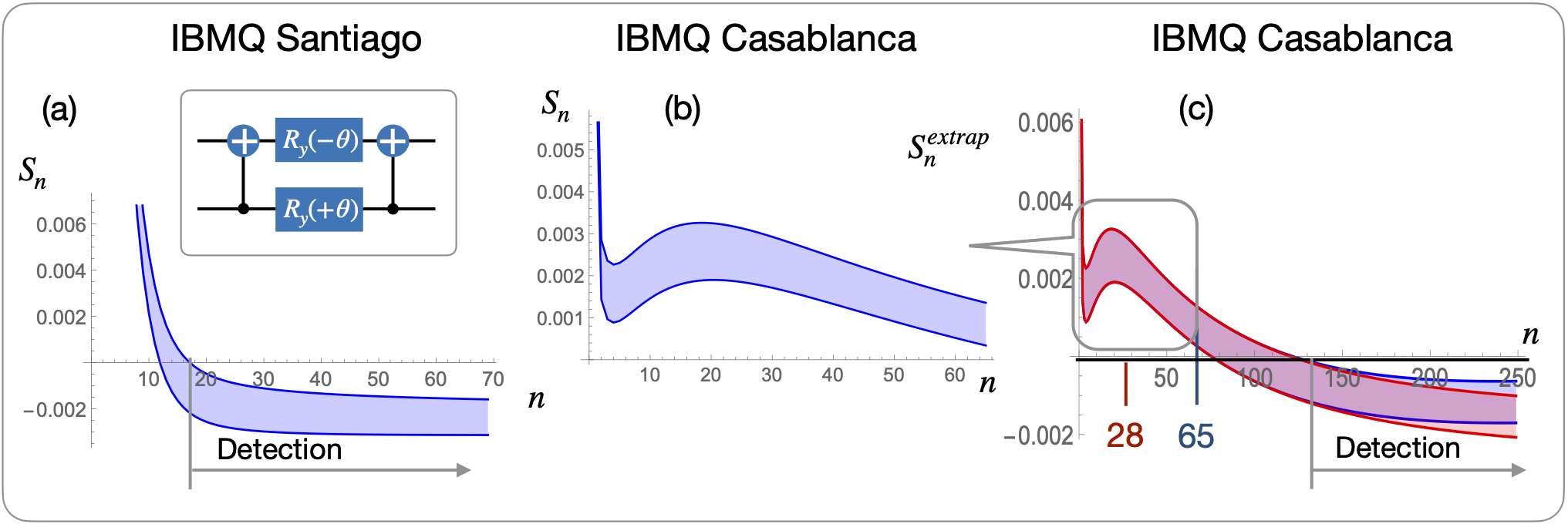}
\par\end{centering}
\caption{\label{fig: int heat leak}Inset: the circuit used for evaluating
the $S_{n}$ inequalities under the intrinsic noise of the IBM processors.
The initial condition is the ground state. $R_{y}$ is a single qubit
rotation around the $y$ axis. (a) For qubits $2$ \& $3$ of the
Santiago processor and $\theta=1.0$, the environment leads to a violation
after $17$ cycles, i.e. $S_{17}<0$. (b) Running the same experiment
on qubits $5$ \& $6$ of the Casablanca processor and $\theta=0.1$,
no violation is observed for the $65$ cycles we were able to run
in this experiment. (c) Using the extrapolation described in Sec.
\ref{subsec:The-scaling-law} based on the favorable $\sqrt{n}$ scaling,
a violation is observed after $130$ cycles. The extrapolation can
be used to detect a heat leak using the first $65$ cycles (blue)
or even just the first $28$ cycles (red).}
\end{figure}

Following the controlled artificial environment demonstration above,
it is interesting to check whether this method can detect the real physical
environment of quantum processors. The following experiment shows that the real environment
is indeed detectable when exploring larger numbers of cycles. The
one-cycle circuit we use is shown in the inset of Fig. \ref{fig: int heat leak}
where $\theta=1$. The value of $\theta$ was chosen arbitrarily and it is not optimal. Other values might detect weaker evolution noise or to observe a violation with fewer cycles. Qubits 1 and 2 of the IBM Santiago processor are
initialized in the state $\ket{00}$. Figure \ref{fig: int heat leak}(a)
shows the values of $S_{n}$ as a function of the number of cycles.
The width of the line corresponds $\pm3\sigma$ uncertainty. Hence,
starting from $17$ cycles, one can see a violation beyond the 3$\sigma$
uncertainty. This result shows that these inequalities can detect
the intrinsic processor noise when using a pure state as the input state. 

Figure \ref{fig: int heat leak}b shows the results of a similar experiment
carried out on the IBM Casablanca processor, for 65 cycles and $\theta=0.1$.
The value of $\theta$ was modified from the previous experiment as
we now wish to demonstrate the utility of the scaling feature presented
in Sec. \ref{subsec:The-scaling-law}. In this case, there is no violation
within the number of cycles measured, but it seems that if more cycles
were measured a violation could have been observed. Fortunately, from
the analysis in Sec. \ref{subsec:The-scaling-law} we know that for
large enough $n$ the contribution of the last cycles decays like a gaussian
in the number of cycles $w_{k}^{(n)}\propto e^{-k^{2}/n}$. Consequently, 
only the first $\xi\sqrt{n}$ terms contribute to the value of $S_{n}$.
$\xi$ is roughly 3 but its exact value depends on the required accuracy of
the clipped sum with respect to the full sum. This suggests the $65$
cycles we have measured can be used to evaluate $S_{n}$ with $n>65$. Setting
$\xi=\pi$ and $M=65$, we can extrapolate up to $\left\lfloor (65/\pi)^{2}\right\rfloor =428$
with an error of $10^{-5}$, which is not observable on the scale
of Fig \ref{fig: int heat leak}c. The blue curve, which overlaps with
the red curve until $n\sim140$, is generated by taking the measured
$\{R_{k}\}_{k=0}^{65}$ and setting $R_{\ensuremath{k}}=0$ for $k\ge66$
in the expression for $S_{n}$. We find negative values that indicate
the presence of evolution noise starting from $n=130$. We conclude
that the noise is detectable with only $65$ cycles. This illustrates
the efficiency of the $S_{n}$ inequalities.

It seems that perhaps even $65$ samples are more than enough for
observing the violation. The red curve is $S_{n}$ with $R_{k}$ data
from only $28$ cycles. The red band shows extrapolation based on
only $28$ points. According to eq. (\ref{eq: erfc}) at $140$ cycles, for
example, the error should be smaller than $1\times10^{-3}$, which
is too big to confirm a violation. Yet, in practice, at $n=140$ the
extrapolation based on $28$ points is almost indistinguishable from
the one exploiting $65$ points. The bound given in eq. (\ref{eq: erfc})
on the truncation error of $S_{n}$ is rather loose since it assumes that the
signs of the coefficients $w_{k}^{(n)}$ do not alternate as a function
of $k$, and that all recurrence probabilities are one. Thus, in practice,
the contribution of the terms with high cycle number to the sum can
be much smaller. In general, it is possible to start with a small
number of cycles, and then add more cycles until convergence is achieved
at the extrapolated point. Potentially, other initial states can lead
to earlier detection of evolution noise, but the goal of this example
is to illustrate the advantage of the $\sqrt{n}$ scaling and not
to find the optimal detection scheme.

\subsection{Noise Detection Beyond the Hot-Environment Limit\label{subsec:Noise-Detection hot}}

\begin{figure}
\includegraphics[width=11cm]{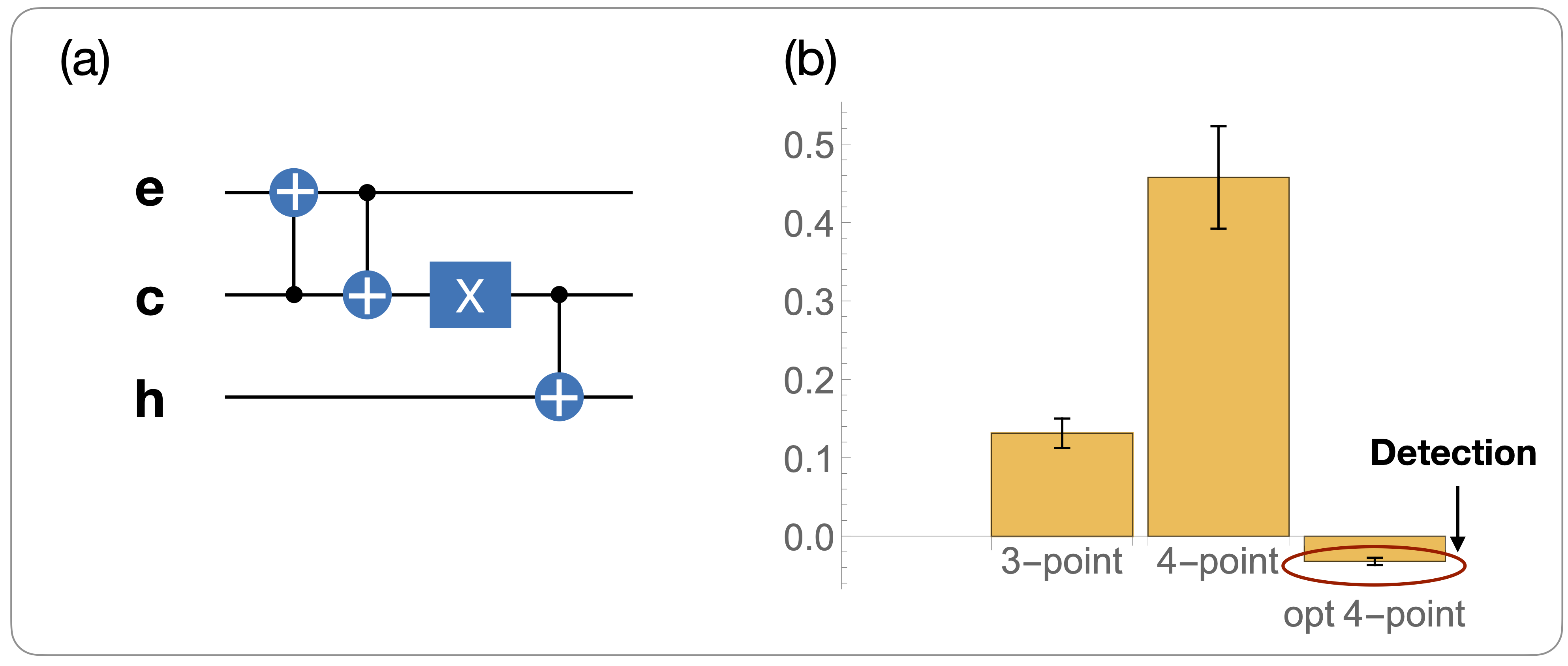}

\caption{\label{fig: hot env}An experiment that confirms the capability of
the periodicity inequalities to detect hot environments which cannot
be detected using two-point passivity-based bounds {[}see eq. (\ref{eq: T undetect}){]}.
Running the circuit (a) for two and three cycles, we observe in (b)
that the $S_{2}$ and $S_{3}$ inequalities are not violated, but
the optimized three-cycle inequality is violated and therefore confirms
the coupling to the hot qubit 'e'. The experiment was done on the
Ourense IBM processor. See main text for details.}
\end{figure}

Evolution noise created by hot environments deserves special attention.
In the extreme case where the environment is fully mixed, i.e. infinite
temperature, the observed system experiences unital dynamics. Unital maps are characterized by the fact that the fully-mixed state is a fixed point of these maps \cite{nielsen2002introduction,smolin2005entanglement}. Decoherence processes are another example of unital map. 
For practical purposes, it is enough that the temperature is significantly
larger than the energies of the system. Although not unitary, unital dynamics
are consistent with second-law-like inequalities \cite{peres2006quantum,esposito2011second,uzdin2018second}
and passivity-based inequalities \cite{bassett1978alternative,uzdin2018global,uzdin2021passivity}.
Thus, unital dynamics cannot violate these inequalities, and therefore
these inequalities cannot be used to detect unital noise. One can
expect that if the environment is slightly less hot so that the system
thermalizes to a state, which slightly deviates from a fully mixed
state, it will still be hard to detect the noise this environment
induces on the system. In Ref. \cite{uzdin2021passivity} a theorem was derived
on the environment temperature $T_{env}$ that can still be detected:
there is no observable $C$ in the Hilbert space of the system that
satisfies 
\begin{eqnarray}
\left\langle C\right\rangle _{final}-\left\langle C\right\rangle _{inial}\ge0,\label{eq: C pasivity like}
\end{eqnarray}
for any unitary or unital transformation, that can yield $\left\langle C\right\rangle _{final}-\left\langle C\right\rangle _{inial}\le0$,
i.e. be violated, when the system is coupled to temperature $T_{env}>T_{\text{undetec}}$
where
\begin{eqnarray}
T_{\text{undetec}}=\frac{\text{max}(E_{\text{env}})-\text{min}(E_{\text{env}})}{\text{min}(\mc B_{n}^{\text{vis}}-\mc B_{n-1}^{\text{vis}})},\label{eq: T undetect}
\end{eqnarray}
$\{E_{\text{env}}\}$ are the energy levels of the environment, and
$\mc B_{n}^{\text{vis}}$ is the $n$-th eigenvalue of $-\ln\rho_{\text{sys}}^{\text{initial}}$
(sorted in an increasing order of their size). Hence, if the temperature
of the environment exceeds $T_{\text{undet}}$, it is undetectable
by inequalities of the form (\ref{eq: C pasivity like}). However,
this bound is based on the one-cycle scheme with just one final
state, in contrast to the multi-cycle approach presented in this paper.
Thus, the periodicity inequalities have the potential to detect this
type of environment. In the following experiment, we use the circuits
shown in Figure \ref{fig: hot env}(a). To generate thermal qubits
for the environment and the system, an ensemble of pure states is
used (see \cite{henao2019experimental}). The inverse temperatures
$\beta_{h}=0.6$, $\beta_{c}=3.5$ and $\beta_{e}=0.5$ are chosen
so that the condition given in eq. (\ref{eq: T undetect}) guarantees
that the environment is too hot to be detected using one-cycle inequalities
on observables. As indicated by the negative value of the 3rd bar
in Figure \ref{fig: hot env}(b), by evaluating the optimized three-cycle
inequality (\ref{eq: opt 4pnt}) the heat leak becomes detectable. 

In summary, this experiment illustrates another added value of
the periodicity inequalities with respect to similar inequalities, and that the optimized 3-cycle bound can detect environments that the $S_{3}$
cannot detect, although both of them use the same experimental data. 

\subsection{\label{subsec:A-three-qubit-gate}A three-qubit gate in trapped ions
and superconducting circuits }

Next, we explore the results of $S_{n}$ for some common gates with simple outputs and consider the regime in which these results would provide benefit as diagnostic.
Consider a gate $U$ such as cnot, Toffoli (ccnot), Fredkin (cswap),
etc. that satisfies $U\ket{\psi_{A}}=\ket{\psi_{B}},U\ket{\psi_{B}}=\ket{\psi_{A}}$
i.e. $U$ is a two-state permutation so that $U^{2}$ is the identity
operator. If the initial condition is $\rho_{0}=\ketbra{\psi_{A}}{\psi_{A}}$
or $\rho_{0}=\ketbra{\psi_{B}}{\psi_{B}}$, the ideal output is $R_{n}=\{1,0,1,0,1,...\}$.
As a result $S_{n}=\sum_{k=0}^{\left\lfloor n/2\right\rfloor }w_{2k}=1/2$.
When the predicted $R_{n}$ is so simple, there are many ways to quantify
the deviation of the device from its expected behavior. For example,
one can use$\text{\ensuremath{\left|R_{n}^{exp}-R_{n}\right|}}$.
However, this quantity is always positive and a large value of $\left|R_{n}^{exp}-R_{n}\right|$
can appear due to coherent error, e.g. miscalibrated gates that
change the intended unitary. Thus, $\left|R_{n}^{exp}-R_{n}\right|$
may not be associated with noise created by an environment.
In contrast, negative values of $S_{n}$ are a clear indication of
evolution noise. 

In our last experiment, we run sequentially the Toffoli gate with the
initial state $\ket{1_{A}1_{B}0_{C}}$ where $A$ and $B$ are the
controlling qubits. The experiment is carried out both on the IBMQ
superconducting processors and on a TIQC. The TIQC results are obtained
on the University of Maryland Trapped Ion (UMDTI) quantum computer,
which is described in \cite{linke2017experimental}. This experiment
is performed using a linear chain of five $^{171}\text{Yb}^{+}$
ions in a room-temperature Paul trap under ultrahigh vacuum with the
qubit encoded in two hyperfine ground states of $\text{Yb}^{+}$.
We re-analyze raw data from results reported in \cite{murali2019full},
where three of the five ions are treated as qubits while the others remain
idle. For more information on the UMDTI hardware, see Appendix IV.

Figure \ref{fig: ion}(a) shows the $S_{n}$ plots for various IBM
processors and the UMDTI as a function of the number of cycles $n$.
The $\left|R_{n}^{exp}-R_{n}\right|$ deviation with respect to the
ideal output is depicted in Figure \ref{fig: ion}(b). While the $\left|R_{n}^{exp}-R_{n}\right|$
fluctuates, the $S_{n}$ curves are monotonically decreasing. Using
tools that are beyond the present paper one can prove that the $S_{n}$
series must monotonically decrease if the evolution is unitary and
periodic \cite{PeriodicityPatent,uzdin2021methods}. On the other hand,
the fluctuations of $\left|R_{n}^{exp}-R_{n}\right|$ and other measures,
which are based on distance from the expected output, may be artificial.
For example, if the noise is caused by spontaneous emission, the initial
state is an excited state, and the ideal recurrence probabilities
are $R_{n}=\{1,0,1,0,..\}$, then for odd $n$ values, the noise
will decrease $\left|R_{n}^{exp}-R_{n}\right|$ as it will
take the system closer to the ground state and reduce the overlap
with the initial excited state. Note that the deviation from $1/2$
in Figure \ref{fig: ion}(b) can arise from a coherent error in the
implementation. It appears that the TIQC data is the closest to the
predicted $1/2$ value. Thus, an increase in the $S_{n}$ plot is
an indication for evolution noise, which is not observed here. 

In our next test, we group pairs of Toffoli gates as a single cycle.
The recurrence probability for this two-Toffoli cycle is $R_{n}'=R_{2n}=\{1,1,1...\}$.
Since $\sum_{k=0}^{n}w_{k}^{(n)}=0$ it follows that for an ideal
evolution $S_{n}'=\sum_{k=0}^{n}w_{k}^{(n)}R_{2k}=\sum_{k=0}^{n}w_{k}^{(n)}=0$
for all n. Figure \ref{fig: ion}(c), shows the experimental values
of $S_{n}'$ for the various quantum processors. The same trends as in Figure 6a are observed. Note that in 6(b) Casablanca is more noisy for n=3 and n=7 but it is the other way around for n=5. We conjecture that the reason there are strong fluctuations in $\left|R_{k}^{exp}-R_{k}\right|$ and even lack of consistency regarding which processor is more noisy is related to coherent errors. These findings indicate that it is advantageous to check the performance of devices in terms of $S_{n}$ when the ideal values of $R_{k}$ are {1,0,1,0,...} or {1,1,1,1,...} compare to direct evaluation of the individual $R_{k}$ values, i.e. $\left|R_{k}^{exp}-R_{k}\right|$.
\begin{figure}
\includegraphics[width=15cm]{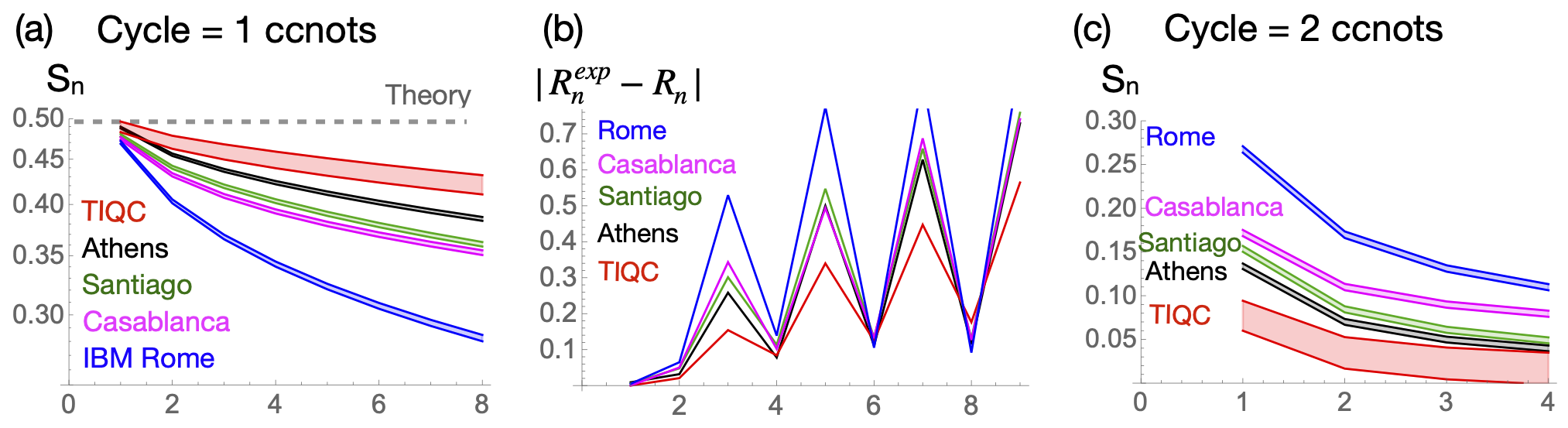}

\caption{\label{fig: ion} When a circuit changes the initial state to an orthogonal
state, and in the next cycle returns to the initial state it holds
that $S_{n}=1/2$. (a) $S_{n}$ values for various quantum processors
where the circuit contains a single Toffoli gate. The plot shows that
the TIQC exceeds the performance of various IBM processors which are
further away from $1/2$. (b) The comparison of the theoretical and
experimental values of the recurrence probabilities, $\left|R_{n}^{exp}-R_{n}\right|$,
seems like a reasonable choice for comparing different processors.
Yet, this quantity strongly fluctuates and in this case,  no
processor is consistently better at all time points. In contrast,
the $S_{n}$ shows a clear and consistent difference between the various
processors. (c) Same as (a), but this time the cycle contains two
consecutive Toffoli gates. In this case, the ideal evolution yields
$S_{n}=0$. Here as well, the TIQC performs better than the tested
IBM superconducting quantum processors.}
\end{figure}

\section*{Concluding remarks}

We have presented a set of multi-cycle inequalities valid for periodically-driven
quantum systems. Presently, we do not claim that these bounds will
mature into a practical method for diagnosing evolution noise in quantum
circuits, but our results present a clear case that this research
direction warrants further investigation. The goal of this paper is
to show that it is possible to formulate constraints that are
customized to periodic driving and that they have an added value with
respect to other known constraints. The periodicity inequalities have
three appealing features. The first is the operational advantage:
they use pure states and only the recurrence probability is measured.
The second is that they treat the circuits as black boxes. The third is the $\sqrt{n}$ scaling
law, which states that only the first $\propto\sqrt{n}$ cycles contribute
to the $S_{n}$ inequality. 

The construction of $S_{n}$ has additional potential beyond the inequalities
studied in this paper. For example, we have shown it is directly related
to purity loss and therefore may have further application beyond detecting
evolution noise, e.g. for devising a simple protocol for entanglement
entropy measurement \cite{PeriodicityPatent,uzdin2021methods}. 

Going beyond digital quantum circuits, our method could find applications in other scenarios with periodic driving, such as the simulation of Floquet many-body systems which have applications in the study of time crystals \cite{else2016floquet} and prethermalization \cite{peng2021floquet}. Typical experimental sequences allow the calculation of bound violations based on existing data, e.g. in \cite{zhang2017observation}.
\\

R.U. is grateful for support from Israel Science Foundation (Grant
No. 2556/20). A.M.G. is supported by a JQI Postdoctoral Fellowship.
N.M.L. acknowledges financial support from NSF grant no. PHY-1430094
to the PFC@JQI, and the Maryland-ARL Quantum Partnership, grant no.
W911NF1920181. 

\section*{Appendix I - The Zero Temperature Problem in Thermodynamic Constraints}

There is a broad family of thermodynamic and thermodynamically-inspired
constraints. Reviewing how exactly the zero temperature problem manifests
in each one is beyond the scope of the present paper. In this appendix,
we give as examples the Jarzynski fluctuation theorem, the second
law in microscopic quantum systems, and passivity-based inequalities. 
he Jarzynski equality in systems whose Hamiltonian returns to
its initial value reads
\[
\langle\langle e^{-\beta W}\rangle\rangle=\sum p_{i}p_{i\to j}e^{-\beta(E_{j}-E_{i})}=1,
\]
where $p_{i\to j}$ is the transition probability generated by the
unitary driving, and $W$ is the work invested in creating this transition.
It is tempting to ignore all input states save the ground state, which
is solely occupied in the $\beta\to\infty$ limit, however, the exponentially
small probabilities are multiplied by an exponentially large factor
(e.g. in the transition to the ground state from some other input
state). Thus, all input states are important in the ultra-cold limit.
This manifests in an exponentially diverging number of shots \cite{jarzynski2006rare}
\[
N=e^{\beta\left\langle W\right\rangle }
,\]
which makes this scheme impractical for pure-state initial conditions.
\subsection*{The second law in microscopic systems and passivity inequalities}

For an isolated system that starts in thermal equilibrium the relevant
statement of the second law is: energy cannot be extracted by applying
a periodic force on the system. The translation to quantum systems
is as follows: given a time independent system Hamiltonian $H_{s}$
the initial density of the system is the Gibbs state
\begin{eqnarray}
\rho_{\beta}=e^{-\beta H_{s}}/tr[e^{-\beta H_{s}}],\label{eq: rho beta}
\end{eqnarray}
where $\beta$ is the inverse temperature. A periodic driving force
refers to a periodic driving Hamiltonian $V(t)$ that satisfies $V(t)=V(\tau)=0$
where $\tau$ is the end of the process. The accumulated effect of the
time dependent Hamiltonian $H=H_{s}+V(t)$ can be described by a unitary
evolution operator, such that the final density matrix is $\rho_{0}=U\rho_{\beta}U^{\dagger}$.
The second law in the case states that the final average energy of
the bare Hamiltonian is equal to or larger than that of the initial
state,
\begin{eqnarray}
\left\langle H_{s}\right\rangle _{f}=tr[\rho_{f}H_{s}]\ge tr[\rho_{\beta}H_{s}]=\left\langle H_{s}\right\rangle _{\beta}.\label{eq: energy passivity}
\end{eqnarray}
This result can be derived in several ways. In particular, it is
a special case of the Clausius-like inequalities for small quantum
systems (\cite{uzdin2018second} and references therein).

When $\beta$ is finite this statement is not trivial because the excited states
are populated according to (\ref{eq: rho beta}). Thus, the system
can be reduce its energy by moving population to lower states. However,
it turns out the unitary transformation (created by the periodic driving)
cannot do that. Non-unitary dynamics, such as cooling can reduce the
energy but this always involves heating up some other object. Thus,
an experimental violation of (\ref{eq: energy passivity}) can be
used to deduce that the system is not sufficiently isolated and interaction
with an unaccounted environment takes place.

The key point in the context of this paper is that when $\beta\to\infty$,
i.e. the ultra-cold limit, for all practical purposes only the ground
state is initially populated. In this case, it does not matter if
the dynamics are unitary or not, and if more objects are involved. The
system is already in the minimum possible energy and therefore it
can only increase. Since (\ref{eq: energy passivity}) cannot be violated
in this case, it cannot be used to detect an environment.

A similar problem exists in passivity-based frameworks. In these framework
the basic inequality stems from the passivity of the operator $B=F(\rho_{0})$, where $\rho_{0}$ is the initial state and $F(x)$ is a monotonically-decreasing function in $x\in[0,1]$. When $\beta\to\infty$, $\rho_{0}\to\ketbra 00$
and therefore the basic passivity inequality
\[
tr[\rho_{f}B]-tr[\rho_{0}B]\ge0,
\]
becomes trivially satisfied for pure states since $B=-\ketbra 00$
(up to a multiplicative positive constant) for any $F$, and the passivity
prediction is
\[
tr[\rho_{f}\ketbra 00]\le1,
\]
which always holds even if the evolution is not unitary. Thus it cannot be used to detect non-unitary environments.
\section*{Appendix II - The Two-Cycle Inequality}

As a basis for comparison, the usual one-cycle inequality, eq. (\ref{eq: F x GP}),
yields $R_{0}\ge R_{1}$ and $R_{0}\ge R_{2}$ which for pure states
is trivial since $R_{0}=1$ and $R_{0},R_{1}\le1$. The inequality
(\ref{eq: 3 point}) can be assigned with different interpretations
via different rearrangements: 
\begin{eqnarray}
R_{2}\ge4R_{1}-3R_{0},\label{eq: 3 point form II}\\
\frac{1}{4}(R_{2}+3R_{0})\ge R_{1},\label{3 point form III}\\
(R_{0}-R_{1})\ge\frac{1}{4}(R_{0}-R_{2})\ge0.\label{3 point form IV}
\end{eqnarray}
Inequality (\ref{eq: 3 point form II}) provides a \emph{lower} bound
on $R_{2}$ (the two-point bound yields an upper bound $R_{2}\le R_{0}$),
while (\ref{3 point form III}) offers a refined upper bound on $R_{1}$.
Since $R_{0}\ge\frac{1}{4}(R_{2}+3R_{0})$ this bound is always tighter
than the two-point bound $R_{0}\ge R_{1}$. Interestingly this bound
is using the information on $R_{2}$, so the two endpoints are used for
bounding the midpoint. Inequality (\ref{3 point form IV}) compares
the change in the recurrence probability in the first half to the
cumulative change $R_{0}-R_{2}$. It suggests that the change cannot
occur just in the second half of the evolution; at least a quarter
must take place in the first half. Similarly one can write $3(R_{0}-R_{1})\ge R_{1}-R_{2}$
and directly compare the two halves. Note however that in this form
the right-hand side might become negative. Another added value of
the form (\ref{3 point form IV}) is that it makes it easier to compare
with the two-point inequalities $R_{0}\ge R_{1}$ and $R_{0}\ge R_{2}$.
The inequality (\ref{3 point form IV}) shows that $R_{0}-R_{1}$
is not just non-negative but also larger than another non-negative
number, $\frac{1}{4}(R_{0}-R_{2})$. Thus, it is tighter than the
two-point prediction $R_{0}-R_{1}\ge0$. 

\section*{Appendix III - Derivation of Equation (\ref{eq: SnA})}

The n-th discrete shifted derivative can be written $(1-A_{+})^{n}\rho_{0}$
where $A_{+}^{n}\rho_{m}=\rho_{m+n}$ and $A_{-}^{n}\rho_{m}=\rho_{m-n}$.
Note, that $A_{+}$ and $A_{-}$ commute. Using this notation $S_{n}$
can written as 
\begin{equation}
S_{n}=\frac{1}{2^{2n}}tr[(1-A_{+})^{n-1}\rho_{0}[(1-A_{+})^{n-1}-A_{+}(1-A_{+})^{n-1}\rho_{0}].
\end{equation}
Next, we use the shift invariance property $tr[\rho_{n}\rho_{m}]=tr[\rho_{0}\rho_{m-n}]$
to obtain
\begin{equation}
S_{n}=\frac{1}{2^{2n}}tr[\rho_{0}(1-A_{-})^{n-1}[(1-A_{+})^{n-1}-A_{+}(1-A_{+})^{n-1}\rho_{0}],
\end{equation}
or 
\begin{equation}
S_{n}=\frac{1}{2^{2n}}tr[\rho_{0}(1-A_{-})^{n-1}(1+A_{-})^{n-1}(1-A_{+})\rho_{0}].\label{eq: Sn expr a}
\end{equation}

Next, we apply the hermitian conjugate of the expression inside the
trace. It still gives $S_{n}$ since it is real but changes $A_{+}$
to $A_{-}$ and vice versa and we get the alternative expression 
\begin{equation}
S_{n}=\frac{1}{2^{2n}}tr[\rho_{0}(1-A_{-})^{n-1}(1+A_{-})^{n-1}(1-A_{-})\rho_{0}].\label{eq: Sn expr b}
\end{equation}
 Combining Eq. (\ref{eq: Sn expr a}) and Eq. (\ref{eq: Sn expr b})
we get

\begin{equation}
S_{n}=\frac{1}{2^{2n}}tr[\rho_{0}(1-A_{-})^{n-1}(1+A_{-})^{n-1}(2-A_{-}-A_{+})\rho_{0}].
\end{equation}

Finally, since $(1-A_{-})^{n}(1-A_{+})^{n}=(2-A_{-}-A_{+})^{n}$ we
get eq. (\ref{eq: SnA})
\begin{align*}
S_{n} & =\frac{1}{2^{2n}}tr[\rho_{0}(2-A_{-}-A_{+})^{n}\rho_{0}]\\
 & =tr[\rho_{0}(\frac{1}{2}-\frac{1}{4}A_{-}-\frac{1}{4}A_{+})^{n}\rho_{0}].
\end{align*}
\section*{Appendix IV - The University of Maryland Trapped-Ion Quantum Computer}

The UMDTI quantum computer is described in \cite{linke2017experimental}.
Briefly, two-photon Raman transitions are used to control the qubit
state, encoded in two magnetic-field-insensitive hyperfine ground
states of $^{171}\text{Yb}^{+}$ ions held in a linear chain in a
Paul trap. Individual manipulation of each qubit is performed by splitting
one of the Raman laser beams into several beams, each controlled by
an independent acousto-optic modulator channel and focused onto a
single ion in the chain. Single-qubit gate operations are executed
by creating laser pulses of controlled phase and duration while two-qubit
gates are compiled from single-qubit gates and a laser-driven entangling
Ising gate (XX or $e^{i\chi\sigma_{x}\sigma_{x}}$) following the
M\o lmer-S\o rensen gate scheme \cite{molmer1999multiparticle,solano1999deterministic,milburn2000ion},
which creates entanglement between pairs of qubits via the shared
harmonic oscillator modes of the ion chain in the trap. These modes
act as an information bus with which the qubits are temporarily entangled.
Modulation of the Raman beam amplitude is used to leave the qubits
disentangled from these motional degrees of freedom at the end of
the gate operation \cite{zhu2006arbitrary,choi2014optimal}.

\bibliographystyle{unsrtnat}
\bibliography{name}

\end{document}